# Infrared resonance-lattice device technology

Robert Magnusson
Department of Electrical Engineering, University of Texas Arlington, Arlington, Texas 76019
Resonant Sensors Incorporated, Arlington, Texas 76010
Tiwaz Technologies LLC, Arlington, Texas 76006

Yeong H. Ko, Kyu J. Lee, Fairooz A. Simlan, Pawarat Bootpakdeetam, Renjie Chen
Department of Electrical Engineering, University of Texas Arlington, Arlington, Texas 76019

Debra Wawro Weidanz, Susanne Gimlin, Soroush Ghaffari
Resonant Sensors Incorporated, Arlington, Texas 76010

## ABSTRACT

We present subwavelength resonant lattices fashioned as nano- and microstructured films as a basis for a host of device concepts. Whereas the canonical physical properties are fully embodied in a one-dimensional periodic lattice, the final device constructs are often patterned in two-dimensionally-modulated films in which case we may refer to them as photonic crystal slabs, metamaterials, or metasurfaces. These surfaces can support lateral modes and localized field signatures with propagative and evanescent diffraction channels critically controlling the response. The governing principle of guided-mode, or lattice, resonance enables diverse spectral expressions such that a single-layer component can behave as a sensor, reflector, filter, or polarizer. This structural sparsity contrasts strongly with the venerable field of multi-layer thin-film optics that is basis for most optical components on the market today. The lattice resonance effect can be exploited in all major spectral regions with appropriate low-loss materials and fabrication resources. In this paper, we highlight resonant device technology and present our work on design, fabrication, and characterization of optical elements operating in the near-IR, mid-IR, and long-wave IR spectral regions. Examples of fabricated and tested devices include biological sensors, high-contrast-ratio polarizers, narrow-band notch filters, and wideband high reflectors.

## 1. INTRODUCTION

The exploration of wave propagation in periodic optical systems represents a longstanding and esteemed discipline within science, engineering, and technology, dating back over a century. One of the most basic examples of such systems is the traditional diffraction grating. Available in various types, including reflection and transmission gratings with straightforward configurations, these gratings can be crafted with remarkable efficiency, often achieving nearly 100% effectiveness in their primary diffraction order. Within strictly periodic gratings or lattices, incident light undergoes a spatially periodic phase transformation, giving rise to diffracted waves traveling in specific directions determined by the wavelength, as governed by the grating equation. This fundamental characteristic underpins their spectroscopic attributes. In their basic form, they serve as crucial commercial components with a myriad of practical applications. A broader perspective encompasses the realm of diffractive optics, offering more flexible design layouts and application possibilities. Diffractive optics, in its entirety, encompasses essential theory, design methodologies, fabrication techniques, and diverse applications. Diffractive optical elements (DOEs) are engineered to regulate various aspects of optical wave behavior, including spatial distribution, spectral characteristics, energy distribution, polarization state, and propagation direction. Common applications include spectral filters, diffractive lenses, antireflection surfaces, beam splitters, beam steering devices, laser mirrors, polarization controllers, beam shapers, and couplers, among others. These components find extensive usage across a broad spectrum of fields, including lasers, fiber-optic communications, spectroscopy, medical technology, integrated optics, imaging, and many other optical systems, covering a wide range of spectral regions [1-5].

*magnusson@uta.edu; phone 1 817 272-2552; www.leakymoderesonance.com

Subwavelength diffractive optics emerges as a specialized subset within the DOE domain [6]. Subwavelength diffraction ensues when the elemental period Λ is smaller than the operational wavelength. Then, only zero-order forward- and backward-diffracted waves propagate with all higher orders being evanescent. At normal incidence θ=0, the operational wavelength satisfies $λ>λ_R=nΛ$ where $λ_R$ is the Rayleigh wavelength and n is the larger of the refractive indices of the substrate and cover regions. Subwavelength DOEs can be engineered using periodic arrays of intricately shaped particles within a homogeneous host medium. Alternatively, periodic profiles can be etched into substrates or films, offering limitless possibilities for customizing profile shapes. These particles and surface structures typically take the form of pillars, blocks, or rods composed of metals, dielectrics, or semiconductors. Within the subwavelength domain, evanescent diffraction orders can interact with leaky Bloch modes, generating a guided-mode resonance response. Background on the physics, formulation, experiments, characterization, and applications of resonant photonic elements can be found in selected references [7-27]. Examples of fabricated optical lattices are presented in Fig. 1.

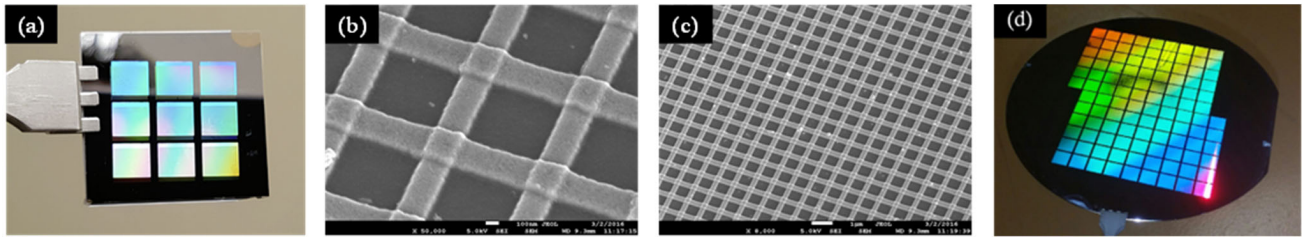

Figure 1. Representative fabricated optical lattices. (a) 3x3 array of 5 mm$^2$ 1D silicon elements. (b) Close-up of a 2D lattice. (c) Broad view of the 2D lattice. (d) 108 1 cm$^2$ guided-mode resonance (GMR) devices on a 6-inch wafer.

In this paper, we address resonant device technology for IR applications. Thus, we exemplify design, fabrication, and characterization of optical elements operating in the near-IR, mid-IR, and long-wave IR spectral regions. We begin by reviewing and explaining the physical principles governing this device class. Then we present representative examples of fabricated and tested devices including biological and environmental sensors, high-contrast-ratio polarizers, narrow-band notch filters, and wideband high reflectors.

## 2. GUIDED-MODE RESONANCE PHYSICS

Figure 2 illustrates a subwavelength resonance system. The incident optical wave undergoes a guided-mode resonance (GMR) on coupling to a leaky eigenmode of the layer system. There may result high-efficiency, or perfect, reflection or transmission depending on design and system parametric details. The fields radiated by the leaky modes in this lattice geometry with a symmetric profile can be in phase or out of phase at the edges of the band [28]. At one edge, there is a zero-phase difference, and hence the radiation is enhanced while at the other edge, there is a π phase difference inhibiting the radiation. Analyzing the second order stop bands with numerical models, Vincent and Neviere explained the existence of a non-leaky edge pertinent to symmetric gratings whereas asymmetric grating profiles yielded leaky radiant modes at both band edges [7]. Consistent with these explanations, in Fig. 2(b) at normal incidence and thus under complete symmetry, we see a fully resonant GMR band edge and a non-resonant symmetry-blocked edge that is offset from the GMR edge. The non-resonant edge is currently referenced as a "bound state in the continuum (BIC)." The term BIC appeared in photonics in 2008 [29] referring to the concept treated in [28]. Much earlier, BICs were proposed in hypothetical quantum systems by von Neumann and Wigner [30]. Being grounded in the conceptual approach provided by Kazarinov and Henry (the KH model), we associated absence of radiation at a band edge with a state having $β_I=0$ where $β = β_R + iβ_I$ is the propagation constant of a leaky mode [31]. In a 2007 publication, the BIC edge was designated as the non-leaky band edge and the GMR edge as the leaky edge [32] with some advantages in clarity and connection with the physical principles at hand. In recent literature, the GMR edge is sometimes referred to as a "bright" state and the BIC edge as a "dark" state [33, 34]. In Fig. 2(a), schematic leaky modes seen as counterpropagating are indicated and driven by a normally-incident wave. Contributing to the versatility of spectral engineering with resonant lattices, these modes can be multiple and take on classic mode shapes, or approximations of those, designated as TE$_0$, TE$_1$ …. as noted in Fig. 2(b). A leaky band belongs to each existing mode and the parametric dynamics of these bands are of great scientific interest [35, 36]. Figure 2(b) further illustrates the spectral variations depending on lattice symmetry, noting the generation of a GMR response at the BIC state as symmetry protection is lifted. Thus, the modal BIC states switch from dark to bright in asymmetric systems.

We emphasize that, in this paper, strictly periodic systems are treated. If the subunits, or particles, forming an array are positioned randomly, the assembly will scatter incident light incoherently with random phasing different optical response. Thus, all major properties observed with periodic assemblies will be lost. This includes perfect resonant reflection and generation of discrete propagating and evanescent diffraction orders. Moreover, the concept of band structure becomes irrelevant. The individual particles will resonate light in a local manner with scattering amplitudes and directional properties depending on materials, particle shape, and their proximity to neighboring particles. Fabry-Perot (FP) resonance occurs under reflections between parallel planes possessing refractive-index discontinuities. This universal effect is typically associated with thin films and thin-film systems and is the basis for an important class of commercial components and devices [37]. Mie resonance occurs by analogous reflections but between nonparallel planes and is generally associated with isolated cylindrical and spherical particles [38]. Mie resonance can be viewed as a generalized FP resonance in arbitrary geometries. Mie scattering is a venerable field of study with attendant eminent scientific scholarship [39]. But Mie scattering is not the effect underlying the physics of the periodic resonant optical lattices in focus here.

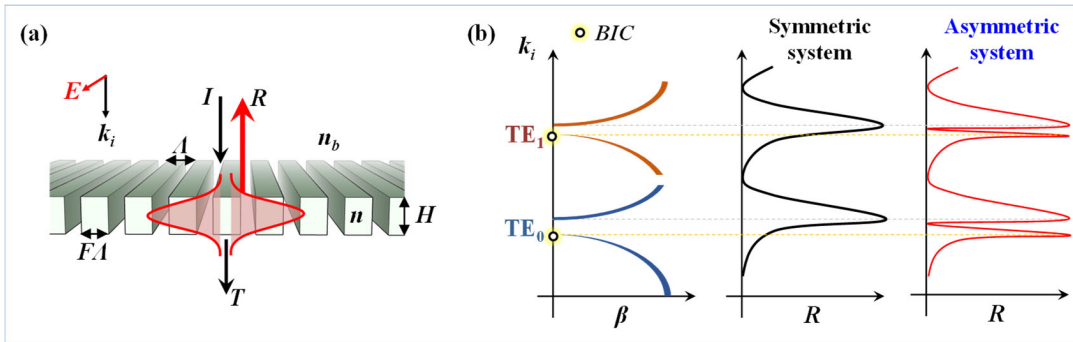

Figure 2. (a) A schematic view of the simplest subwavelength resonance system. The model lattice has thickness ($H$), fill factor ($F$), period ($\Lambda$), and refractive indices of background and lattice material ($n_b$, $n$). When phase matching occurs between evanescent diffraction orders and a waveguide mode, a guided-mode resonance occurs. $I$, $R$, and $T$ denote the incident wave with wavelength λ, reflectance, and transmittance, respectively. (b) A schematic dispersion diagram of a resonant lattice at the second stop band. For a symmetric lattice, the leaky edge supports guided-mode resonant radiation while the non-leaky edge hosts a non-radiant bound state. This picture applies to both TE (electric field vector normal to the plane of incidence and pointing along the grating grooves) and TM (magnetic field vector normal to the plane of incidence) polarization states. Here, the grating vector has magnitude $K = 2\pi/\Lambda$, $k_i = 2\pi/\lambda$, and $\beta$ denotes a propagation constant of a leaky mode.

## 3. GUIDED-MODE RESONANCE SENSORS

Light-based sensors find wide deployment in practical applications. Example high-value fields include medical diagnostics, biomarker discovery, drug development, food safety, industrial process control, and environmental monitoring. The guided-mode resonance sensor operates according to the principles set forth in Fig. 2. The resonance frequency of a guided-mode resonance device changes if structural parameters vary. In biomolecular binding to the sensor surface, an attaching biolayer alters the effective thickness of the resonant layer affecting the resonance wavelength. This is shown schematically in Fig. 3(a). The wavelength change can be monitored with a spectrum analyzer in real time to quantify the binding dynamics. Because a resonant lattice responds to both TE and TM polarization states and at differing spectral points, polarization diversity exists as noted in Fig. 3(b). Moreover, since the lattice can host multiple simultaneous Bloch modes, a modal diversity exists as also shown in Fig. 3(b). These properties enable multiparametric biomolecular analysis [40, 41]. A wide variety of sensor geometries, materials, and system architectures can be implemented. For practical use, the sensor is optimized for sensitivity, configuration, and cost [42]. The GMR sensor has an interesting technical and legal history [43-49].

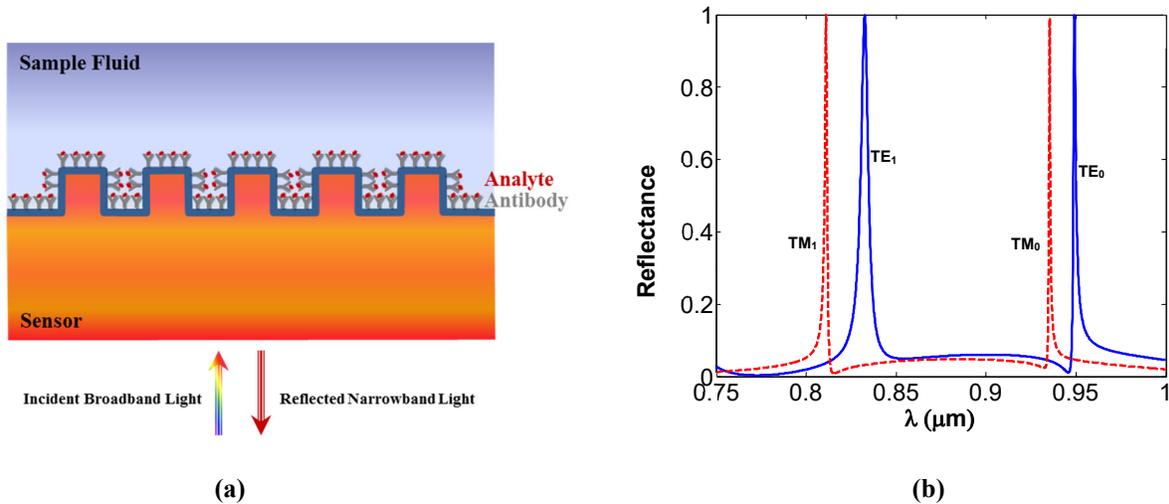

Figure 3. Characteristics of GMR sensors. (a) A guided-mode resonance sensor schematic indicating a binding event. (b) Numerical resonance spectra for a representative dielectric GMR sensor operating in an aqueous environment. Two resonance peaks associated with the lowest waveguide modes for each polarization state appear within the chosen spectral range as noted.

The biosensors used in this research are fabricated using low-cost submicron molding methods. We employ polymers imprinted with submicron grating patterns (~500 nm grating periods, ~100 nm grating height) coated with a high-index dielectric material (such as $TiO_2$) to realize resonant sensors. The sensor measurement methodology schematized in Fig. 3(a) applies where a broadband light source illuminates the GMR biosensors and where a specific wavelength of light is reflected as per Fig. 3(b). These sensors are designed to operate in the near-IR wavelength range (700–900 nm), where most biochemical materials have minimal absorption [42]. The sensor plates used are supplied by Resonant Sensors Incorporated (RSI, Arlington, TX, USA). The ResoSens bioassay system, which includes a light source, a temperature control module, a spectrometer, and a mobile stage comprises a reader. The Bionetic microarray plates applied here are manufactured by RSI [50].

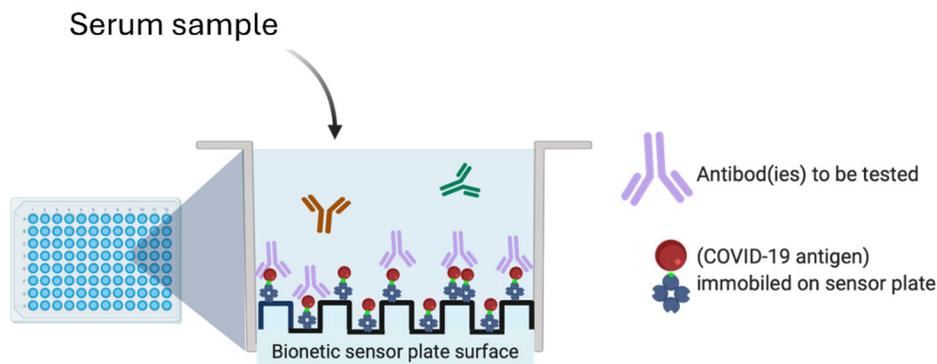

Figure 4. Schematic rendition of a 96-well RSI sensor plate with biochemical activity as noted for a COVID antigen and antibody reaction. Image created with BioRender.com.

Figure 4 indicates an RSI plate containing 96 microwells in standard format. The GMR sensor plate resides under the microwell assembly with the sensing surface towards the well containing the solution under test. COVID-19 antigen(s) such as receptor binding domain spike protein (RBD) are linked to the sensor surface via linker molecules. The bioreaction begins as the antibodies bind to the antigen and change the effective surface properties yielding a GMR wavelength shift that grows relative to the quantity of physical binding occurring. The complete dynamic reaction is thus quantified which contrasts strongly to endpoint-only data. Example results are provided in Fig. 5 denoting the resonance spectral shifts (in pm) for a confirmed COVID-19 positive patient serum sample containing antibodies against the RBD antigen, and a confirmed negative patient serum sample with no antibodies reacting to the RBD antigen. Results are measured in quadruplicate and averaged. A reference buffer is subtracted from the data. There is clear differentiation between the positive and negative cases with full reaction dynamics recorded for each patient sample.

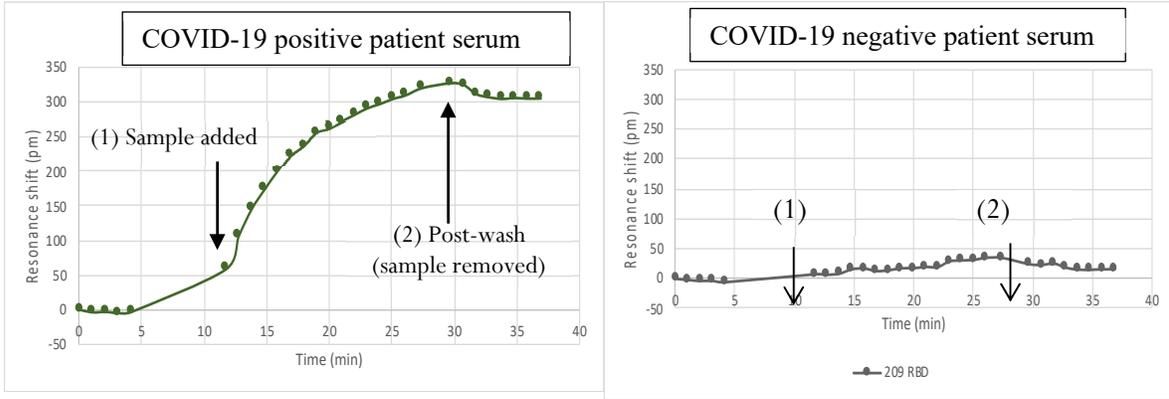

Figure 5. GMR sensor data relevant to COVID-19 antigen-antibody reaction. COVID-19 positive patient serum samples containing antibodies that react with RBD antigen immobilized on the sensor surface provide a measurable resonance shift, while negative serum samples show minimal shift.

## 4. GUIDED-MODE RESONANCE POLARIZERS

As noted above, the fundamental GMR effect is polarization sensitive. This motivates search for applications in polarization discrimination. Numerous past publications pertain to this effort [51, 52]. Recently, we have designed and fabricated compact low-loss, ultra-high extinction ratio polarizers based on multilayer resonant periodic lattices [53]. The building block of the cascaded device is an individual periodic lattice, or metasurface in common terminology, polarizer containing subwavelength periodic patterns of crystalline silicon on top of a quartz substrate. These metasurfaces are fundamentally GMR grids. Cascading all-dielectric low-loss individual polarizers offers high-quality performance in a compact format as compared to conventional polarizers as we confirm by rigorous computations. The design possesses an appropriate angular tolerance of the transmission spectra for both TE and TM polarization states.

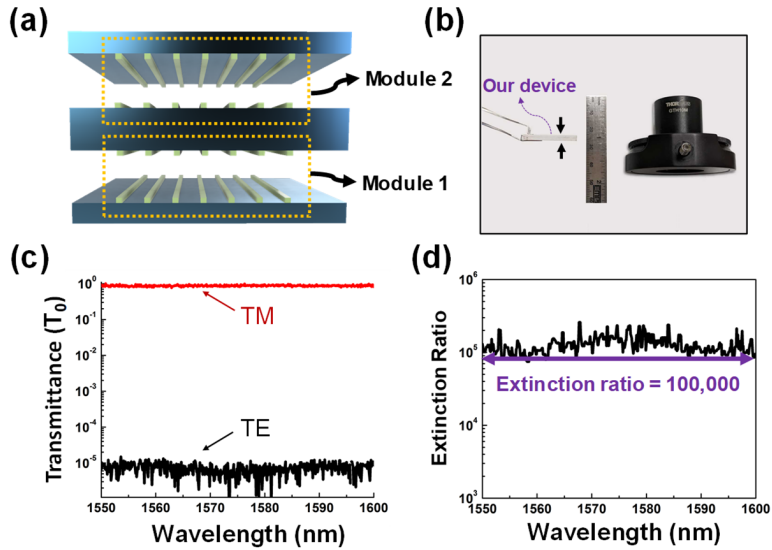

Figure 6. A fabricated stacked multi-metasurface polarizer. (a) Schematics of a device containing two dual-metasurface polarizer modules. (b) Photograph of the polarizer and Thorlab's Glan-Thompson polarizer. (c) Log-scale transmission spectra for TE and TM polarization states. (d) Extinction ratio of the fabricated polarizer [53].

## 5. GUIDED-MODE RESONANCE FILTERS

The filtering capabilities of the basic GMR effect have been known for decades and widely reported [10-18]. The early research pertained mostly to the visible and near-IR spectral regions. Recently, results on LWIR device fabrication have appeared [54, 55]. Here, we briefly exemplify our experimental work in demonstrating high-quality LWIR lattice resonance devices. The LWIR spectral region spanning ~8 to 12 μm is useful for a host of scientific and industrial applications. As traditional multilayer film components are not straightforwardly realized at these bands, alternate solutions are of interest. Hence, there is motivation to apply the GMR concept as a new method to achieve for devices and components in this band.

Accordingly, we provide design, fabrication, and testing of bandstop filters based on the guided-mode resonance effect. Focusing on the zero-contrast grating architecture, we successfully fabricate prototype filters in the Ge-on-ZnSe or Ge-on-ZnS materials system. Thus, Fig. 7(a) shows a top view of the fabricated GMR filter obtained with a scanning electron microscope (SEM), indicating a straight and uniform Ge-based line grating structure. Figure 7(b) shows atomic force microscope (AFM) images of the GMR filter. With these images, the optical lattice parameter set is measured as ($\Lambda$ = 2.85 μm, F = 0.65, $d_g$ = 0.535 μm, $d_h$ = 1 μm). The F is somewhat larger than in the original device design due to experimental inaccuracies in the photolithography and dry etching processing. Figure 7(c) shows the experimental and theoretical (using measured parameters) spectral response of the GMR filter for both TE and TM polarizations. The spectra of the fabricated GMR filter were measured using a Nicolet iN10 Fourier transform infrared (FTIR) spectrometer with iZ10 transmission measurement attachment. A 10,000:1 extinction ratio wire-grid linear polarizer was used to set the polarization state. To quantify the spectral response of the resonant grating, the measured spectrum was normalized by the intensity of the input beam. As seen in these curves, TE and TM peaks are located at 9.6 μm and 8.64 μm with 0.42 μm and 0.657 μm FWHM (full width at half maximum) respectively, with high transmission sidebands. In the sidebands, $T_0$ is degraded by a significant back side (substrate) reflection of R~17%. In view of this reflection loss, not compensated in the experimental data, the experimental curves and the theoretical spectra are in good agreement, which indicates the device dimensions obtained by SEM and AFM measurements are accurate. There results appear in [56].

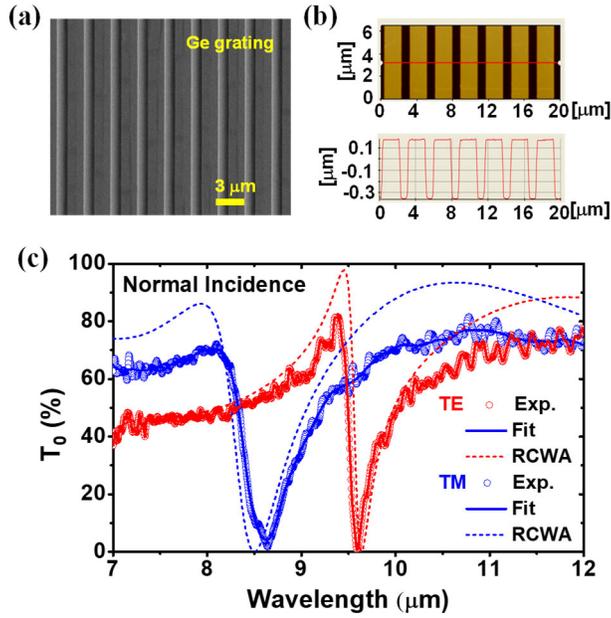

Figure 7. Fabrication and measurements of the 1D optical lattice device. (a) SEM and (b) AFM images of the fabricated 1D Ge-based grating. In this experiment, the measured grating parameters set is ($\Lambda$ = 2.85 μm, F = 0.65, $d_g$ = 0.535 μm, $d_h$ = 1 μm). The refractive index of deposited Ge is 4.1. (c) Measured and calculated zero-order transmission spectra are compared. For the IR region, the spectrum is measured by FTIR and polarizer. Due to substrate back-side reflection (R~17%), the measured transmittance falls somewhat below the theoretical quantity [56].

## 6. GUIDED-MODE RESONANCE REFLECTORS

There are numerous examples of well-designed and fabricated GMR reflectors in the literature where most results pertain to visible and near-IR wavelengths. For reasons mentioned in Section 5, there is interest in extending the technology to cover longer wavelengths for use in terrestrial imaging, spectroscopic applications, night-vision systems, and medical and industrial laser technologies. Thus, we are developing LWIR reflectors based on 1D and 2D gratings fabricated in Ge films on ZnSe or ZnS substrates. The refractive indices of these materials are spectrally dispersive such that $n_c(\lambda) = n(\lambda) + ik(\lambda)$. In the 5-20 µm band, they are commercially available as low-loss dielectric materials such that $k(\lambda)$ is approximately zero [57]. Figure 8(a) shows a 2D-patterned Ge/ZnSe device design with a corresponding spectral map in Fig. 8(b) with dark red color denoting full reflection. Due to the homogeneous sublayer, the grating-with-sublayer reflectors outperform comparable grating-only reflectors such that the spectra generated are significantly more tolerant of parametric deviations than spectra of gratings without the sublayer as quantitatively shown in [23]. If fill factors $F_x=F_y$, the device will be polarization independent at normal incidence and will have a significant angular aperture (~±10°) across which it remains approximately polarization independent.

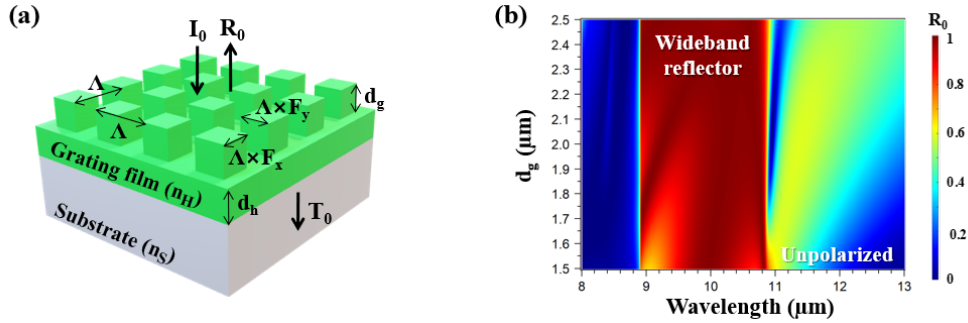

Figure 8. (a) Schematic of a 2D grating-with-sublayer design for polarization independent reflectors. (b) Calculated $R_0$ map as a function of grating depth ($d_g$). Parameters are $\Lambda$=3.71 µm, $F_x=F_y$ = 0.67, $d_g$=variable, and $d_h$= 0.72 µm.

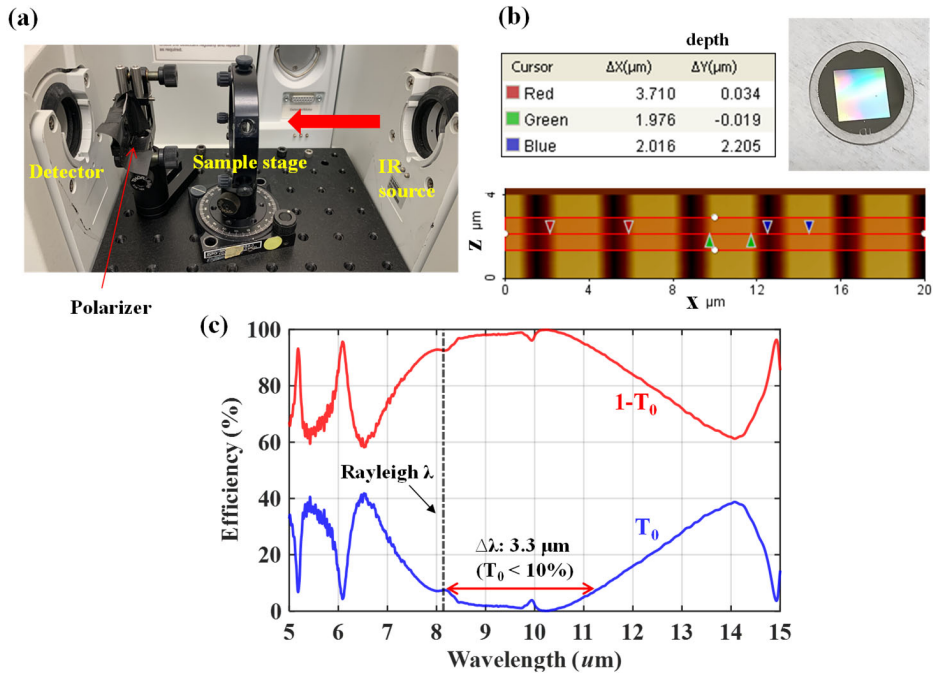

Figure 9. Experimental results pertaining to development of LWIR wideband reflectors. (a) Photograph of the FTIR measurement setup. (b) AFM image characterizing the grating structure along with a photograph of the actual device residing on a 1-inch substrate. (c) Measured transmission spectrum of the fabricated device.

Results on LWIR reflector performance are shown in Fig. 9. The measurement setup, depicted in Fig. 9(a), utilizes an FTIR spectrometer along with a polarizer to assess the reflectance performance of the sample under TM polarization incidence. Figure 9(b) illustrates a clear square region of the Ge grating, while atomic force microscope images reveal 1D grating structures with specific parameters (period = 3.71 µm, fill factor = 0.53, grating depth = 2.20 µm). Figure 9(c) shows the measured spectra of the fabricated sample, with the dashed line denoting the Rayleigh wavelength. By measuring the zeroth-order transmittance ($T_0$), the reflectance spectrum is estimated as $1-T_0$ under the assumption of lossless materials for both the Ge film and ZnS substrate. The achieved spectra demonstrate a wide bandwidth of $\Delta\lambda$= 3.3 µm across which reflection levels $R_0 \sim 1-T_0 > 90\%$.

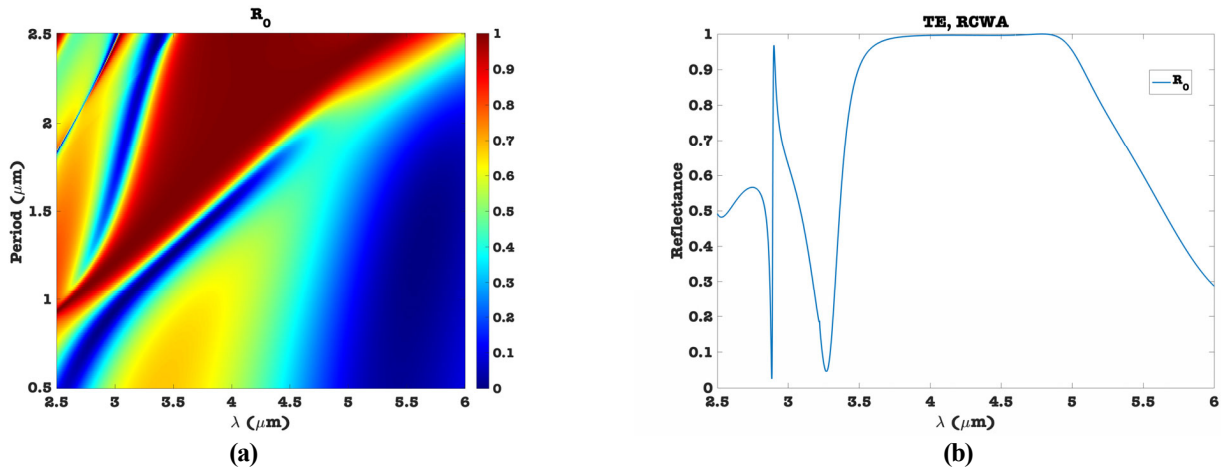

Figure 10. (a) Mid-IR spectral map for a Ge/glass resonant mirror. Parameters are $d_g$= 0.75 µm, $d_h$= 0.16 µm, F = 0.3, $n_{Ge}$=4.1, $n_s$ =1.4, $k_s$=0. (b) Cut at period $\Lambda$=2.3 µm.

Finally, another key atmospheric window spans ~ 3-5 µm in wavelength. It is of interest to investigate feasibility and design of GMR devices in this region as well. In this spectral range, glass has minimal loss with k~0 whereas in the 8-12 µm it is highly lossy and unusable for substrates. Figure 10(a) shows a spectral map relative to period for a resonant optical lattice on a glass substrate revealing high-reflectance loci with significant bandwidths. Figure 10(b) is a spectral cut at $\Lambda$=2.3 µm. We note that relative to Zn-based substrate media, glass is a particularly economic option.

## 7. CONCLUSIONS

The subject of this paper is resonant device technology for IR applications. Representative resonant optical elements operating in the near-IR, mid-IR, and long-wave IR spectral regions are discussed. Whereas the principles behind these devices have been known for many years, robust and repeatable fabrication methods came much later and are still under development. We began the article by reviewing and explaining the physical basis governing this device class. This physics requires excitation of lateral Bloch modes by evanescent-wave coupling with concomitant existence of a leaky band structure. For each supported mode there is a leaky band edge with guided-mode resonance radiation and a non-leaky edge for the bound-state dark channel. For the periodic lattices in play, there are no dominant effects from simple Fabry-Perot resonances in lattice ridges even though this mechanism is claimed in 100s of published papers. Moreover, particle-based Mie scattering does not cause resonance reflection or any other important observed effects even though this is asserted in 1000s of articles. The erroneous explanations of resonance lattice physics and corresponding spectral expressions appear in many journals published by SPIE, Optica (OSA), AIP and numerous other outfits including Science and Nature.

Proceeding to example applications, we present the GMR sensor. Pertinent high-sensitivity, label-free biosensing methodology has major practical applicability due to efficient and economic manufacturability of the basic sensing elements. We review our work in realizing high-extinction polarizers that are interesting on account of their extreme material sparsity and good performance albeit working across relatively limited spectral extents. We conclude with presentation of LWIR optical component examples of narrow-band filters and wideband reflectors. We show designs and example spectral data found by successful photolithographic processes that we have developed. With the use of durable materials and methods, our work provides a practical foundation for realizing a variety of functional devices operating in the LWIR spectral region with sparse film structures that can be relatively expeditiously deposited. These include bandpass filters, band-stop filters, resonant sensors, and polarizers that have favorable size, weight, and power properties with potential extensions to spatially aperiodic, or chirped, imaging sensors, metalenses, detectors, and other device concepts.

## ACKNOWLEDGEMENTS

This research was supported, in part, by the UT System Texas Nanoelectronics Research Superiority Award funded by the State of Texas Emerging Technology Fund as well as by the Texas Instruments Distinguished University Chair in Nanoelectronics endowment. Additional support was provided by the National Science Foundation under Award No. ECCS-1809143. Tiwaz Technologies LLC acknowledges support by NSF via SBIR award No. 2304394. Bionetic sensor plates and all reagents were provided by Resonant Sensors Incorporated. Parts of this research were conducted in the UT Arlington Shimadzu Institute Nanotechnology Research Center.